\documentclass[12pt]{article}
\usepackage{graphics}

\def\hybrid{\topmargin -20pt    \oddsidemargin 0pt
        \headheight 0pt \headsep 0pt
        \textwidth 6.25in       
        \textheight 9.5in       
        \marginparwidth .875in
        \parskip 5pt plus 1pt   \jot = 1.5ex}

\hybrid

\def\baselinestretch{1.2}

\catcode`\@=11

\def\marginnote#1{}
\newcount\hour
\newcount\minute
\newtoks\amorpm
\hour=\time\divide\hour by60
\minute=\time{\multiply\hour by60 \global\advance\minute by-\hour}
\edef\standardtime{{\ifnum\hour<12 \global\amorpm={am}%
        \else\global\amorpm={pm}\advance\hour by-12 \fi
        \ifnum\hour=0 \hour=12 \fi
        \number\hour:\ifnum\minute<10 0\fi\number\minute\the\amorpm}}
\edef\militarytime{\number\hour:\ifnum\minute<10 0\fi\number\minute}

\def\draftlabel#1{{\@bsphack\if@filesw {\let\thepage\relax
   \xdef\@gtempa{\write\@auxout{\string
      \newlabel{#1}{{\@currentlabel}{\thepage}}}}}\@gtempa
   \if@nobreak \ifvmode\nobreak\fi\fi\fi\@esphack}
        \gdef\@eqnlabel{#1}}
\def\@eqnlabel{}
\def\@vacuum{}
\def\draftmarginnote#1{\marginpar{\raggedright\scriptsize\tt#1}}

\def\draft{\oddsidemargin -.5truein
        \def\@oddfoot{\sl preliminary draft \hfil
        \rm\thepage\hfil\sl\today\quad\militarytime}
        \let\@evenfoot\@oddfoot \overfullrule 3pt
        \let\label=\draftlabel
        \let\marginnote=\draftmarginnote
   \def\@eqnnum{(\theequation)\rlap{\kern\marginparsep\tt\@eqnlabel}%
\global\let\@eqnlabel\@vacuum}  }

\def\preprint{\twocolumn\sloppy\flushbottom\parindent 2em
        \leftmargini 2em\leftmarginv .5em\leftmarginvi .5em
        \oddsidemargin -.5in    \evensidemargin -.5in
        \columnsep .4in \footheight 0pt
        \textwidth 10.in        \topmargin  -.4in
        \headheight 12pt \topskip .4in
        \textheight 6.9in \footskip 0pt
        \def\@oddhead{\thepage\hfil\addtocounter{page}{1}\thepage}
        \let\@evenhead\@oddhead \def\@oddfoot{} \def\@evenfoot{} }

\def\numberbysection{\@addtoreset{equation}{section}
        \def\theequation{\thesection.\arabic{equation}}}

\def\underline#1{\relax\ifmmode\@@underline#1\else
        $\@@underline{\hbox{#1}}$\relax\fi}

\def\titlepage{\@restonecolfalse\if@twocolumn\@restonecoltrue\onecolumn
     \else \newpage \fi \thispagestyle{empty}\c@page\z@
        \def\thefootnote{\fnsymbol{footnote}} }

\def\endtitlepage{\if@restonecol\twocolumn \else \newpage \fi
        \def\thefootnote{\arabic{footnote}}
        \setcounter{footnote}{0}}  

\catcode`@=12
\relax

\def\figcap{\section*{Figure Captions\markboth
        {FIGURECAPTIONS}{FIGURECAPTIONS}}\list
        {Figure \arabic{enumi}:\hfill}{\settowidth\labelwidth{Figure
999:}
        \leftmargin\labelwidth
        \advance\leftmargin\labelsep\usecounter{enumi}}}
 \relax
\def\tablecap{\section*{Table Captions\markboth
        {TABLECAPTIONS}{TABLECAPTIONS}}\list
        {Table \arabic{enumi}:\hfill}{\settowidth\labelwidth{Table
999:}
        \leftmargin\labelwidth
        \advance\leftmargin\labelsep\usecounter{enumi}}}
 \relax
\def\reflist{\section*{References\markboth
        {REFLIST}{REFLIST}}\list
        {[\arabic{enumi}]\hfill}{\settowidth\labelwidth{[999]}
        \leftmargin\labelwidth
        \advance\leftmargin\labelsep\usecounter{enumi}}}
 \relax
\makeatletter
\newcounter{pubctr}
\def\publist{\@ifnextchar[{\@publist}{\@@publist}}
\def\@publist[#1]{\list
        {[\arabic{pubctr}]\hfill}{\settowidth\labelwidth{[999]}
        \leftmargin\labelwidth
        \advance\leftmargin\labelsep
        \@nmbrlisttrue\def\@listctr{pubctr}
        \setcounter{pubctr}{#1}\addtocounter{pubctr}{-1}}}
\def\@@publist{\list
        {[\arabic{pubctr}]\hfill}{\settowidth\labelwidth{[999]}
        \leftmargin\labelwidth
        \advance\leftmargin\labelsep
        \@nmbrlisttrue\def\@listctr{pubctr}}}
 \relax
\makeatother
\newskip\humongous \humongous=0pt plus 1000pt minus 1000pt

\newif\ifdtup

\relax

\def\be{\begin{equation}}
\def\ee{\end{equation}}
\def\ba{\begin{eqnarray}}
\def\ea{\end{eqnarray}}

\def\del{\partial}

\def\b{\beta}

\def\e{\epsilon}

\def\m{\mu}

\def\om{\omega}

\def\vphi{\varphi}

\def\no{\noindent}

\def\qq{\qquad}

\def\IR{\relax{\rm I\kern-.18em R}}

\def \ha {{1\over 2}}

\def \ov {\over}

\def\IR{\relax{\rm I\kern-.18em R}}
\def\inv{^{\raise.15ex\hbox{${\scriptscriptstyle -}$}\kern-.05em 1}}

\def\tL{{\tilde L}}

\def\wt{\widetilde W}

\begin{document}

\renewcommand{\theequation}{\thesection.\arabic{equation}}

\newcommand{\beq}{\begin{equation}}
\newcommand{\eeq}[1]{\label{#1}\end{equation}}
\newcommand{\ber}{\begin{eqnarray}}
\newcommand{\eer}[1]{\label{#1}\end{eqnarray}}
\newcommand{\eqn}[1]{(\ref{#1})}
\begin{titlepage}
\begin{center}

\hfill NEIP-00-020\\
\vskip -.05 cm
\hfill hep--th/0011203\\
\vskip -.05 cm
\hfill November 2000\\

\vskip .6in

{\large \bf Supersymmetry and finite-temperature strings
\footnote{Expanded version of talks given by K.S. in the 
{\it 8th International Conference on Supersymmetries in Physics}, 
CERN, Geneva, Switzerland 26 June-1 July 2000, in 
{\it The Ninth Marcel Grossmann Meeting} (to appear in the proccedings), 
Rome, July 2-8, 2000 and  in the conference on {\it Particle Physics and 
Gravitation: Quantum Fields and Strings}, Kolymbari, Greece, 
9-15 September 2000.}}

\vskip 0.5in

{\bf I. Bakas${}^1$ },\phantom{x}{\bf A. Bilal${}^2$ },\phantom{x} 
{\bf J.-P. Derendinger${}^2$ }\phantom{x}and\phantom{x} 
{\bf K. Sfetsos${}^2$}
\vskip 0.1in
{\em ${}^1\!$Department of Physics, University of Patras \\
GR-26500 Patras, Greece\\
\footnotesize{\tt bakas@ajax.physics.upatras.gr}}
\vskip .2in
{\em ${}^2\!$Institut de Physique, Universit\'e de Neuch\^atel\\
  rue Breguet 1, CH-2000 Neuch\^atel, Switzerland\\
\footnotesize{\tt 
adel.bilal,jean-pierre.derendinger,konstadinos.sfetsos@unine.ch}} 

\vskip .2in

\end{center}

\vskip .7in

\centerline{\bf Abstract}

\no
We describe finite temperature $N=4$ superstrings in $D=5$ by an
effective four-dimensional supergravity of the thermal winding modes
that can become tachyonic and trigger the instabilities at the
Hagedorn temperature. Using a domain-wall ansatz, exact solutions to
special BPS-type first order equations are found. They preserve
half of the supersymmetries, contrary to the standard perturbative
superstring at finite temperature that breaks all supersymmetries.
Our solutions show no indication of any tachyonic instability and
provide evidence for a new BPS phase of finite temperature
superstrings that is stable for all temperatures. This would 
have important consequences for a stringy description
of the early universe.

\vskip .4 cm
\noindent
\end{titlepage}
\vfill
\eject

\def\baselinestretch{1.2}
\baselineskip 16 pt
\noindent

\def\tT{{\tilde T}}
\def\tg{{\tilde g}}
\def\tL{{\tilde L}}


\section{Inroduction}

A $d$-dimensional field theory at finite temperature $T={1\over \b}$ 
is formulated as a theory with a Euclidean time taking values on a 
circle of circumference $\b$. There can then be no time dependence 
any more, in agreement with temperature being  only well-defined in a 
time-independent equilibrium. Bosonic fields must be periodic around 
the circle while fermionic ones must be antiperiodic \cite{KAPU}.
This leads to different Fourier modes, i.e. to different thermal 
masses for bosons and fermions. Thus, if one starts with a 
supersymmetric theory, introducing finite temperature  breaks the 
supersymmetry.

In (closed) superstring theory things are similar but more subtle. 
Whenever a dimension is compactified, the closed string can wind an 
arbitrary number of times around this circle. At zero temperature, 
when an ordinary spatial dimension is compactified the spectrum of 
the superstring is modified by the appearance of corresponding 
momentum and winding states in a modular invariant way. These give 
a positive contribution to the masses of the corresponding states. 
At finite temperature however, we must recover the different 
boundary conditions for fermions and bosons around the thermal 
circle. Modular invariance then fixes all signs for the contributions 
of the different spin structures to the partition function \cite{aw}.
This results in a GSO projection different from the usual one at zero 
temperature. In particular, at zero temperature the state with no 
oscillator excitations (in the RNS formalism) would be a tachyon and 
is eliminated by the GSO projection. At finite $T$ however, such a 
state, with unit winding number, is kept by the modified GSO projection 
and its mass squared is $-8 \alpha' +{\b^2\over \pi^2}$. For large 
$\b$ (small $T$) this is positive, but it turns negative at 
$T=T_H\equiv {1\over \pi \sqrt{8\alpha'}}$, called the Hagedorn 
temperature. Thus at high temperature certain winding modes turn 
tachyonic, signaling an instability and a transition to a new phase 
where the corresponding field has acquired a vacuum expectation value.

Usually the existence of this instability for $T\ge T_H$ is 
interpreted as due to the non-convergence of the thermal partition 
function $Z={\rm Tr} e^{-\b H}$ because the number of string states 
contributing to the $n^{\rm th}$ level grows like $e^{c\sqrt{n}}$. 
This reason looks very different from the above argument but one 
should realize that in string theory modular invariance relates 
this exponential growth at high energies to the behavior of the 
low-lying momentum and winding modes.

Since one can view the thermal instability as solely due to certain 
winding modes becoming tachyonic, it should be possible to get an 
accurate description of the physics near the critical temperature 
by studying an effective theory of these modes (fields). In 
refs. \cite{ak,adk} this effective theory was found for all known 
five-dimensional $N=4$ superstring theories at finite temperature, 
which effectively are four-dimensional. A relevant subset of 
potentially tachyonic winding modes was isolated from the 
non-perturbative BPS mass formula, and an effective supergravity 
theory for these modes, coupled to the usual dilaton and temperature 
modulus, was constructed. This theory is briefly reviewed in 
section 2. The supergravity potential clearly exhibits domains 
where tachyonic modes are present.

Rather than looking at perturbation theory of this supergravity 
(which one could do directly in the full string theory), we try to 
find exact solutions \cite{bbds,sfet}. To do so, we 
make a domain-wall ansatz for the metric and use a certain technique 
(more or less well-known from recent studies of RG-flows in the AdS/CFT 
correspondence)
to reduce the second order equations to
first order ones. This is reminiscent of BPS conditions and, indeed, 
we have proven \cite{bbds} that all solutions of the first order 
equations preserve half of the supersymmetries, i.e. they are really 
BPS. This will be described in section 3. 

The full set of first order equations is still too complicated to be 
solved exactly, but certain consistent truncations lead to various 
type II or heterotic theories. The solutions in each of 
these cases have been described in great detail in ref. \cite{bbds} and 
also in ref. \cite{sfet}. Here, in section 4, we describe only their main 
properties. A striking feature is that almost all of these solutions 
interpolate between low and high temperature, typically including 
regions of strong coupling. 
They exhibit no sign of instability or phase transition whatsoever. 
This apparent stability fits well with their BPS property.

The usual instability of strings as the Hagedorn temperature $T_H$ 
is reached is obtained from a perturbative analysis where supersymmetry is 
completely broken by the finite temperature. Instead, we have found 
solutions which preserve half of the supersymmetries. 
We are tempted to conclude that they correspond to a new, more stable 
phase of superstrings (or at least of the effective supergravity) 
which does not undergo any phase transition as the temperature is 
raised. This may have important consequences in a stringy description 
of the early universe.

\section{Review of the Effective Thermal Supergravity}

One starts with the six-dimensional $N=4$ closed superstring theories 
(heterotic on $T^4$ or type II on $K3$), compactifies one dimension 
on a circle of radius $R_6$ and the Wick-rotated Euclidean time on the 
thermal circle of radius $R=(2\pi T)^{-1}$. As discussed above, the GSO 
projection has to be changed, resulting in a modified thermal mass 
formula. It is enough to restrict the attention on those modes that 
become tachyonic first as the temperature is raised. These are certain 
dyonic modes with winding numbers $\pm 1$. There are three such $\pm$ 
modes. Which pair of them turns tachyonic first depends on the values of the 
moduli $s = {\sqrt{2}/ g_{{\rm het}}^2}$, $ t = {\sqrt{2}RR_6/  
{\alpha}_{{\rm het}}^{\prime}}$ and $u = {\sqrt{2}R/ R_6}$
(the four-dimensional 
gravitational coupling constant is normalized as $\kappa=\sqrt{2}$
and ${\alpha}_{{\rm het}}^{\prime} = 4 s$),
which determine whether one is in the heterotic, type IIA or type IIB 
sector. Altogether, one has to find an effective supergravity for the 
$3\times 2$ dyonic modes and the 3 moduli $s,t,u$. 
The restriction to these modes only 
truncates the theory from $N=4$ to an $N=1$. This provides 
enough constraints to determine the (bosonic part of the) effective 
theory (see ref. \cite{adk} for details and ref. \cite{bbds} for a shorter 
review). It is of the general form 
\be
S=\int {\rm d}^4 x\ \sqrt{G} \left( {1 \over 4}R 
- {1 \over 2} K_{I\bar{J}}\del^\m\Phi^I\del_\m\bar{\Phi}^{J} 
- V(\Phi, \bar{\Phi}) \right) \ ,  
\label{sugralagr}
\ee
where $\Phi^I$, $I=1,2,\dots ,9$ denotes the bosonic parts of the chiral 
superfields and $K_{I\bar{J}}$ is the K\"ahler metric derived from 
a K\"ahler potential $K$. The potential $V$ is given by 
$V= {1\over4}\, e^K \left( K^{I\bar{J}} W_{;I} \overline{W}_{;\bar{J}} 
- 3 W \overline{W} \right)$ where 
$W_{;I} = {\partial W \over \partial \Phi^I} 
+ {\del K \over \del \Phi^I} W$ and $W$ is the superpotential. 
The potential and the superpotential were given in ref. \cite{adk}.

The analysis of the thermal mass spectrum in ref. \cite{adk} reveals that 
the directions in which tachyonic instabilities could occur are 
the real directions. It is thus enough to restrict oneselves to 
this case, where we have a total of 9 real fields.
Writing $s=e^{-2 \phi_1}, t=e^{-2\phi_2}, u=e^{-2\phi_3}$ 
and denoting the real parts of the truncated dyonic modes as 
$z^\pm_a = {\rm Re}\,Z_a^\pm $ one finds 
($x_\pm^2\equiv \sum_{b=1}^3 (z^\pm_b)^2$)
\be
S=\int {\rm d}^4 x\ \sqrt{G} \left(
{1 \over 4}R - \ha \sum_{i=1}^3 (\del_\mu\phi_i)^2
-  \sum_{a=1}^{3} { (\del_\m z^+_a)^2\over (1 - x_+^2)^2} 
-  \sum_{a=1}^{3} { (\del_\m z^-_a)^2\over (1 - x_-^2)^2} 
- V \right) \ .
\label{laagra}
\ee
All we need to know about the potential $V$ as a function of the real fields 
is that it can be expressed as
\be
V={1 \over 4} 
\sum_{i=1}^3 \left({\del\wt\over\del\phi_i}\right)^2 
+{1\over 8}\sum_{a=1}^3 (1-x_+^2)^2 \left({\del\wt\over\del z^+_a}\right)^2 
+{1\over 8}\sum_{a=1}^3 (1-x_-^2)^2 \left({\del\wt\over\del z^-_a}\right)^2 
- {3\ov 4} {\wt}^2 \ ,
\label{dina1}
\ee
where the real superpotential $\wt$ is given in terms of the superpotential 
$W$ as $\wt = e^{K/2} W \mid_{{\rm real ~  directions}}$. In
the present case 
\be 
\wt ={1 \over 2} e^{\phi_1+\phi_2+\phi_3} 
- 2 e^{\phi_1} {\rm sinh} (\phi_2+\phi_3) H^+_1 H^-_1
+ e^{-\phi_1}\left(e^{\phi_2-\phi_3} H^+_2 H^-_2
+ e^{\phi_3-\phi_2} H^+_3 H^-_3 \right)
\label{supereal} 
\ee 
where $H^\pm_a\equiv {z^\pm_a\over 1-x_\pm^2}$. 
The form of the theory (\ref{laagra})-(\ref{supereal}) which is obtained
when the $+$ and $-$ winding modes are restricted to behave 
in exactly the same way, i.e. $z_a^+=z^-_a$, 
was given in ref. \cite{bbds}, whereas the more general form 
that is presented here can be found in ref. \cite{sfet}.
In the rest of this contribution we restrict our discussion for simplicity
to the case of $z_a^+=z_a^-\equiv z_a$ following ref. \cite{bbds} while
for its generalization to the case with  
$z_a^+\neq z_a^-$ we refer the reader to ref. \cite{sfet}.

\section{First Order BPS Equations and Supersymmetry}

We make the domain-wall ansatz for the four-dimensional (Euclidean) 
metric:
\be
ds^2 = dr^2 + e^{2 A(r)} \left( dx_1^2+dx_2^2+dx_3^2\right) 
\label{mee2}
\ee
and suppose that all scalar fields only depend on the single coordinate 
$r$. It is then straightforward to verify that the second order 
equations of motion for the metric and the scalar fields $\vphi^I$ 
appearing in eq.(\ref{laagra}) ($\vphi^I= \phi_i$ or $z_a$) are satisfied 
if the $\vphi^I(r)$ and the conformal factor $A(r)$ are solutions of 
the first order equations
\be 
{d\vphi^I\ov dr} = \pm {1\ov \sqrt{2}} K^{I\overline J} 
{\del\wt\over\del\vphi^J}\ ,
\qq {dA\ov dr} =\mp {1\over \sqrt{2}} \wt\ ~.
\label{sugrafoe}
\ee
Such first order gradient equations exist quite generally for the 
domain-wall ansatz if the potential $V$ can be expressed in terms 
of a real superpotential $\wt$ as in (\ref{dina1}). Given the relation 
between $V, \wt$ and $W$, this only depends on the real 
directions and a certain reality property of the K\"ahler potential,
which are realized in a broad class of theories.

One can now study the supersymmetry transformations of the fermions to check
whether a given bosonic solution $\vphi$ breaks or preserves the 
supersymmetries. For example, one has for the lef-handed fermionic component 
$\chi_L^I$ associated with $\vphi^I$:
\be
\delta\chi^I_L = {1\over\sqrt2}\gamma^\mu\,
(\del_\mu\vphi^I)\epsilon_R -
\ha\epsilon_L\, e^{K/2} K^{I\overline J} W_{; J} \ .
\label{chivar}
\ee
Since only $\del_r\vphi^I$ is non-vanishing and 
$e^{K/2} K^{I\overline J} W_{; J} 
= K^{I\overline J}{\del\wt\over\del\vphi_J}
=\pm \sqrt{2} \del_r\vphi^I$, we have
\be
\delta\chi^I_L 
= {1\over\sqrt2} \left( \gamma^r\e_R \mp \e_L\right) \del_r \vphi^I
= {1\over\sqrt2} \del_r \vphi^I P_L (\gamma^r\mp 1) \e \ .
\label{chivartwo} 
\ee
Then, since $\ha  (1\mp \gamma^r)$ is a projector, half of the 
components of $\e$ give $\delta\chi_L^I=0$, i.e. the solutions 
$\vphi^I$ preserve half of the supersymmetries: they are BPS. 
Of course, one also has to check that the supersymmetry variation of the 
gravitino vanishes. This is indeed the case provided 
$\e(r)=e^{A(r)/2} \e_0$, see ref. \cite{bbds} for more details.

The explicit set of six non-linear first-order ordinary differential 
equations for $\phi_1, \phi_2, \phi_3$ and $z_1, z_2, z_3$ is easily 
obtained from (\ref{sugrafoe}) 
using the prepotential (\ref{supereal}) and the K\"ahler 
metric following from (\ref{laagra}). We will 
not write down the eqs. in detail, as they can be found in ref. \cite{bbds}.
We also note that the considerations of this section apply equally well 
to a general four-dimensional $N=1$ supergravity theory coupled to any number
of chiral multiplets.

\section{Truncations to Specific Sectors}

\underline{Type {\rm II}:} We 
further truncate the system to be able to find exact solutions: 
Letting $z_1=z_2=0$, 
$z_3\equiv {\rm tanh}\left({\omega \over 2}\right)$, 
$\phi_1=\phi_2\equiv \phi/\sqrt{2}$ and 
$\phi_3\equiv \chi$ leads to type IIB, while 
letting $z_1=z_3=0$, 
$z_2\equiv {\rm tanh}\left({\omega \over 2}\right)$, 
$\phi_1=\phi_3\equiv \phi/\sqrt{2}$ and 
$\phi_2\equiv \chi$ leads to type IIA.
In both cases one obtains the same $\wt$ and the same equations when 
expressed in terms of $\phi$, $\chi$ and $\omega$. They can be solved 
exactly as ($c$ is a constant of integration)
\be
e^{-2\sqrt{2} \phi}  =  2 \cosh^2\om\ (\ln \coth^2\om +c) -2\ ,
\qq
e^{2 \chi}  = \cosh^2 \om\ e^{\sqrt{2}\phi}\ ,
\label{chiII}
\ee
with $e^{-2\sqrt{2} \phi}\sim T^{-2}$ and $e^\chi\sim g_{\rm II}$ being 
the string coupling. The metric is 
\be
ds^2 ={8 e^{\sqrt{2} \phi}\ov \sinh^2\om \cosh^4\om} d\om^2 
+{e^{-\sqrt{2}\phi}\ov \cosh^2\om} (dx_1^2+dx_2^2+dx_3^2) \ .
\label{metII}
\ee
with $r$ and $\omega$ related by 
$dr = - {2 \sqrt{2} e^{\phi/\sqrt{2}}\ov \sinh\om \cosh^2\om}d\om$. 
One can see from the potential $V$ that $\om$ is tachyonic if 
$e^{-2\sqrt{2} \phi}<4$ at $\om=0$. However, it is obvious from 
the explicit solution (\ref{chiII}) that, independently of the 
value of $c$, one has $T^{-2}\sim e^{-2\sqrt{2} \phi} 
\sim 2 \ln \om^{-2} \to\infty$ as $\om\to 0$. Thus the winding mode 
$\om$ does not become tachyonic. Also, in all cases, the solution 
interpolates from weak to strong coupling ($e^{2\chi}\to\infty$).

We also note that a different type of truncation 
having $z_1=0$, 
$z_2=\pm z_3\equiv {1\over \sqrt{2}} {\rm tanh}{\om\over 2}$, 
$\phi_2=\phi_3\equiv \phi/ \sqrt{2}$, 
$\phi_1\equiv \chi$, leads to a self-dual type II or hybrid 
type II theory. This was presented in great detail in refs. \cite{bbds,sfet}
and will not be repeated here.

\medskip
\noindent
\underline{Heterotic sector:} Finally,
the heterotic sector is obtained by the truncation
$z_2=z_3=0$, $z_1\equiv {\rm tanh}{\om\over 2}$, 
$\phi_2=\phi_3\equiv \phi/ \sqrt{2}$ and $\phi_1\equiv \chi$.
Unfortunately, we have not been able to solve the corresponding 
three equations in closed form. In ref. \cite{bbds} a detailed study 
of various asymptotic regimes, as well as in the vicinity of a critical point 
of the differential equations has been performed, yielding a  
coherent global picture of how the solutions behave. There is one 
weakly coupled region  $\phi\to -\infty,\ \om\to 0$ and three strongly
coupled ones;
region 1: $\phi\to \infty$ and  $\om\to 0$,
region 2: $\phi\to \infty$ and  $\om\to \infty$,
region 3: $\phi\to -\infty$ and  $\om\to \infty$,
as well as a special critical point for $\phi=0$ and $\om=\ln(\sqrt{2}+1)$.
There are solutions going from the special critical point to any of the 
3 strong coupling or the weak coupling regions, as well as solutions
extending  
from weak coupling to regions 1 or 3 and solutions from region 2 to 1 or 3.
We refer the reader to ref. \cite{bbds} for an in depth discussion of these 
solutions.


One expects that certain physical criteria discriminate between 
the various solutions, in particular between different ranges of 
the integration constants. For example, one can look at the metric 
and study the propagation of a test particle (wave) in this 
supergravity background. In general the metrics exhibit certain 
singularities which 
may or may not be admissible. We have found \cite{bbds} that they 
are admissible if the integration constant 
$c\le 0$ for the type II case. For the 
heterotic case, solutions that contain region 3 are not, while 
all others are admissible. It is less clear which solutions 
are admissible when studying the string, rather 
than the particle propagation in these backgrounds.
Another interesting question is the stability of solutions 
against small fluctuations, which is expected for supersymmetric 
solutions. This is currently under study.

\section{Conclusions and speculations}

We have described finite temperature strings in $D=5$ by an effective 
four-dimensional supergravity. Using a domain-wall ansatz, exact 
solutions to special BPS-type first order equations have been found. 
They preserve half of the supersymmetries, contrary to the standard 
perturbative solution at finite temperature that breaks all 
supersymmetries. 
They show no indication of any 
tachyonic instability, since the domain of $(\phi,\om)$ or $(T,\om)$ 
where this instability could occur is avoided by these solutions. 
If this behaviour extends beyond the effective supergravity to the full string 
theory, these solutions will describe a new BPS phase of finite 
temperature superstrings that is stable for all $T$. 

An issue concerning our supersymmetric solutions is whether 
they suffer from Jeans instabilities, typical in thermodynamical systems 
that contain gravity, as the example of hot flat space studied in 
ref. \cite{gross}.
One could be tempted to conclude that such
instabilities cannot be avoided in our solutions especially due to the fact
that they do not support small volumes and according to general arguments
they should collapse into black holes \cite{aw}.
However, the counter argument is that the supersymmetry
of our solutions will ensure their quantum stability and that
no gravitational collapse will occur.
We note at this point that 
the analysis of ref. \cite{gross} was done around hot flat space
and supersymmetry was not even an issue. Our spaces have very different
asymptotics and we think that the conclusions of ref. \cite{gross} are
not directly applicable to our cases.
 
If we want to look for fingerprints of string theory in observational 
data, a good place to study will be the early universe, and 
whatever relics may be observed today. A stringy description of the 
early universe certainly should include finite 
temperature and actually high temperature effects. 
It is conceivable that the differences found 
between a stable BPS high temperature phase and a perturbative phase 
with broken supersymmetry may lead to important changes that eventually will 
be  observed and tested. We leave these exciting questions for further 
study.

\section*{Acknowledgments}

K.S. thanks the organizers of the 9th Marcel Grossmann Meeting,
the SUSY2K, and the Euresco conference 
in Kolymbari, Greece for the opportunities to present our results.
This research was supported
by the Swiss National Science Foundation, the European Union under
contract HPRN-CT-2000-00122, by the Swiss Office 
for Education and Science.


\end{document}